\documentclass[12pt]{article}

\usepackage{graphicx}
\usepackage{dcolumn}
\usepackage{multirow}
\usepackage{bm}
\usepackage{times}
\usepackage{color}
\usepackage{amstext}
\usepackage{amsmath}
\usepackage{amsfonts}

\newcommand{\LNO}{$\rm LaNiO_3$}

\newcommand{\LAO}{$\rm LaAlO_3$}

\newcommand{\SLN}{$\rm LaNiO_3\ ( $$N$$\rm \ u.c.)| LaAlO_3\ ($$N$$\rm \ u.c.)$}
\newcommand{\LSAO}{$\rm LaSrAlO_4\ (001)$}
\newcommand{\STO}{$\rm SrTiO_3\ (001)$}
\usepackage{scicite}
\usepackage{amssymb}

\usepackage{times}
\usepackage{graphicx}
\usepackage{bm}

\newenvironment{sciabstract}{%
\begin{quote} \bf}
{\end{quote}}

\newcounter{lastnote}

\title{Dimensionality Control of Electronic Phase Transitions in Nickel-Oxide Superlattices}
\author
{A. V. Boris$^{1,\S,\ast}$, Y. Matiks$^1$, E. Benckiser$^1$, A. Frano$^1$, P. Popovich$^1$, V. Hinkov$^1$, \\ P. Wochner$^2$,  M. Castro-Colin$^2$, E. Detemple$^2$, V. K. Malik$^3$, C. Bernhard$^3$,\\ T. Prokscha$^4$, A. Suter$^4$, Z. Salman$^4$, E. Morenzoni$^4$,\\ G. Cristiani$^1$, H.-U. Habermeier$^1$, and B. Keimer$^{1\star}$
\\
\normalsize{$^{1}$ Max-Planck-Institut f\"{u}r Festk\"{o}rperforschung,}\\
\normalsize{Heisenbergstrasse 1, D-70569 Stuttgart, Germany}\\
\normalsize{$^{2}$ Max-Planck-Institut f\"{u}r Metallforschung,}\\
\normalsize{Heisenbergstrasse 3, D-70569 Stuttgart, Germany}\\
\normalsize{$^{3}$ Department of Physics, University of Fribourg and}\\
\normalsize{Fribourg Center for Nano Materials, CH-1700 Fribourg, Switzerland}\\
\normalsize{$^{4}$ Laboratory for Muon Spin Spectroscopy, PSI, CH-5232 Villigen PSI, Switzerland}\\
\normalsize{$^{\S}$Department of Physics, Loughborough University,}\\
\normalsize{Loughborough, LE11 3TU, United Kingdom}\\
\normalsize{$^\ast$
E-mail:  A.Boris@fkf.mpg.de.},\normalsize{$^\star$
E-mail:  B.Keimer@fkf.mpg.de.}\\
}
\date{}
\begin{document}

\baselineskip18pt
\maketitle

\begin{sciabstract}
The competition between collective quantum phases in materials with strongly correlated electrons depends sensitively on the dimensionality of the electron system, which is difficult to control by standard solid-state chemistry. We have fabricated superlattices of the paramagnetic metal $\bf LaNiO_3$ and the wide-gap insulator $\bf LaAlO_3$ with atomically precise layer sequences. Using optical ellipsometry and low-energy muon spin rotation,  superlattices with $\bf LaNiO_3$ as thin as two unit cells are shown to undergo a sequence of collective metal-insulator and antiferromagnetic transitions as a function of decreasing temperature, whereas samples with thicker $\bf LaNiO_3$ layers remain metallic and paramagnetic at all temperatures. Metal-oxide superlattices thus allow control of the dimensionality and collective phase behavior of correlated-electron systems.
\end{sciabstract}

\newpage
\baselineskip22pt
Since the discovery of high-temperature superconductivity two decades ago, much effort has been undertaken to explore and understand the quantum physics of strongly correlated electrons in transition metal oxides (TMOs) \cite{Dagotto}. The electronic phases can exhibit radically different physical properties, and a new generation of electronic devices will become possible if the competition between these phases can be systematically controlled \cite{Mannhart}. However, the control options offered by conventional solid-state chemistry are limited. The charge carrier concentration in a TMO compound, for instance, can be modified by chemical substitution \cite{Imada}, but only at the expense of altering the local lattice structure and electronic energy levels in an uncontrolled manner. The dimensionality of the electron system, D, is another key control parameter, because low-dimensional metals are known to be more susceptible to collective ordering phenomena (including spin- and charge-ordering instabilities as well as unconventional superconductivity) than their higher-dimensional counterparts. Some level of dimensionality control has been achieved by synthesizing compounds in the Ruddlesden-Popper series of perovskite structures, which comprise $N$ consecutive TMO layers per unit cell. In principle, the dimensionality of the electron system in these materials can thus be tuned from $\rm D=2$ to 3 by increasing $N$. In practice, however, the synthesis requirements become rapidly more demanding for large $N$, and many Ruddlesden-Popper phases have turned out to be unstable.

Recent advances in the synthesis of TMO heterostructures with atomically sharp interfaces indicate an alternative route towards control of correlated-electron systems \cite{Mannhart}. In principle, the carrier concentration in a heterostructure can be tuned by a gate voltage in a field-effect arrangement, without introducing substitutional disorder, and the dimensionality can be modified by means of the deposition sequence of electronically active and inactive TMO layers. In practice, however, attempts to implement this approach have faced many of the same difficulties encountered in the chemical synthesis of bulk materials. For instance, defects created by interdiffusion or strain relaxation can influence the transport properties of the interfacial electron system in an uncontrolled manner. These difficulties are compounded by the paucity of experimental methods capable of probing the collective phase behavior of electrons in TMO heterostructures. Whereas ferromagnetism and ferroelectricity can be detected based on the macroscopic magnetic or electric field distribution, the identification of two of the most common collective ordering phenomena of correlated electrons, namely charge order and antiferromagnetism, in TMO heterostructures and superlattices is much more difficult.

Motivated by the desire to overcome these difficulties and to realize the potential of TMO heterostructures in controlling collective quantum phases, we have carried out a comprehensive experimental study of superlattices based on the correlated metal LaNiO$_3$ in which the dimensionality of the electron system was used as a control parameter, but the influence of epitaxial strain and defects was carefully monitored. An extensive body of prior work on bulk nickelates provides an excellent background for our study. While bulk $\rm LaNiO_3$ is a 3D Fermi liquid \cite{Eguchi} that remains paramagnetic and metallic at all temperatures, other lanthanide nickelates $R$NiO$_3$ with smaller electronic bandwidths exhibit collective metal-insulator transitions with decreasing temperature \cite{Imada}. In the insulating low-temperature phase, they exhibit a periodic superstructure of the valence-electron charge and a non-collinear antiferromagnetic ordering pattern of spins on the Ni atoms \cite{Mazin,Munoz,Staub1,Staub2}. This implies that the itinerant conduction electrons of $\rm LaNiO_3$ are highly correlated and on the verge of localization. Experiments on a controlled number of atomically thin $\rm LaNiO_3$ layers separated by the electronically inactive wide-gap insulator  $\rm LaAlO_3$ are thus well suited for attempts to control the phase behavior of a correlated-electron system via its dimensionality. We have used wide-band spectroscopic ellipsometry to accurately determine the dynamical electrical conductivity and permittivity, which (in contrast to the $dc$ conductivity) are not influenced by misfit dislocations. Low-energy muons, which are stopped in the SL before they reach the substrate, served as a sensitive probe of the internal magnetic field distribution. Two consecutive, sharp phase transitions in the charge and spin sector revealed by this experimental approach demonstrate that the electronic properties of our SLs are determined by electron correlations, and not by interfacial disorder. By changing the LaNiO$_3$ layer thickness, we demonstrate full dimensionality control over the collective phase behavior.

The superlattices (SLs) were grown by pulsed-laser deposition \cite{huh,SOM} and comprised $N$ consecutive layers of $\rm LaNiO_3$ and the $\rm LaAlO_3$. In order to discriminate between the influence of dimensionality and epitaxial strain, we have grown SLs on both $\rm SrTiO_3$, which induces tensile strain in the overlayer, and $\rm LaSrAlO_4$, which induces compressive strain. Figure 1 shows contour maps of the diffracted X-ray intensity distribution in the vicinity of the $103$ perovskite Bragg peak for three representative samples: $N = 4$ and $N = 2$ SLs grown on \LSAO, and an $N = 2$ SL on \STO. Both the position and the shape of the overlayer reflection are strongly affected by inversion of the type of substrate-induced strain (Figs. 1B and 1C), but remain essentially unchanged by varying the individual layer thicknesses $N$ (Figs. 1A and 1B). A detailed analysis of the substrate-induced strain and relaxation effects is provided in the Online Supplement \cite{SOM}. In the following we show that the transport and magnetic properties of the SLs are only weakly influenced by the strain-induced local structural distortions and interfacial defects, but qualitatively transformed by varying the number of consecutive unit cells within the $\rm LaNiO_3$ layers.

The charge transport properties of the SLs were determined by spectral ellipsometry, which yields the frequency-dependent complex dielectric function, $\varepsilon (\omega)=\varepsilon_1(\omega)+ i \varepsilon_2(\omega)$, related to the optical conductivity $\sigma(\omega)$ by $\varepsilon (\omega) = 1 + 4\pi i\sigma(\omega)/ \omega$. This method is very sensitive to thin-film properties due to the oblique incidence of light, and insensitive to the influence of strain-induced extended defects on the current flow through the atomically thin layers \cite{SOM,CROCMO}. Figures 2A and 2B show the infrared spectra of $\varepsilon_2(\omega)$ for $N = 4$ and $2$ SLs grown on $\rm LaSrAlO_4$ and $\rm SrTiO_3$, respectively, which are representative of the in-plane dielectric response of the metallic $\rm LaNiO_3$ layers. The insets show the corresponding temperature dependencies of $\varepsilon_2$ at a fixed photon energy $\hbar \omega = 30$ meV. The gradual evolution of $\varepsilon_
 2$ with temperature over the far-infrared range confirms that the $N=4$ SLs remain metallic at all temperatures. The $N=2$ SLs, on the other hand, show clear evidence of a metal-insulator transition upon cooling, with a sharp onset at $T_{MI} = 150$ K and 100 K for SLs grown on $\rm LaSrAlO_4$ and $\rm SrTiO_3$, respectively. For $T \gtrsim T_{MI}$, the infrared $\varepsilon (\omega)$ spectrum of $N = 2$ SL is well described by a broad Drude response $\varepsilon(\omega)= \varepsilon_\infty - \omega_{pl}^2/(\omega^2+i\omega\gamma)$ with a ratio of scattering rate and plasma frequency $\gamma / \omega_{pl} \approx 0.2$ (lower shaded line in Fig. 2A) that is typical for bulk complex oxides. The effective mass enhancement $m^{*}/m $ is estimated from the plasma frequency as $m^{*}/m = \frac{4 \pi e^2 n_{Ni}}{m \omega_{pl}^2}$, where $n_{Ni}=\frac{1}{2}\times 1.7\times 10^{22}$ cm$^{-3}$, by assuming one electron per Ni atom. With $\omega_{pl}\approx 1.1$ eV [Fig. 2A and Fig. S7 in the SOM \cite{SOM}] we obtain $m^{*}/m \approx 10$ which is in good agreement with the value for bulk $\rm LaNiO_3$ obtained from specific heat measurements \cite{Xu}. Using the Fermi energy $E_F=0.5$ eV  derived from the thermopower of $\rm LaNiO_3$ \cite{Xu} and $\gamma $ from the Drude model fit to the infrared spectra, we estimate the mean free path as \  $l=\frac{1}{2\pi c\gamma}\sqrt{2E_F/m^*}$.
For the $N = 2$ and $N = 4$ SLs on both substrates we obtain $l = 5-6$ \AA \ and $10-12$ \AA, respectively \cite{SOM}. Remarkably, the mean free path correlates with the individual $\rm LaNiO_3$ layer thickness, testifying to the high quality of the interfaces.

The charge-carrier localization at lower temperature can be readily identified through a rapid drop in $\varepsilon_2(T)$ and progressive deviation of $\varepsilon_2 (\omega)$ from the Drude function due to the formation of a charge gap. The temperature evolution of the real part of the dielectric function provides complementary information about the optical spectral-weight redistribution at $T_{MI}$. Figures 2C and 2D show the temperature dependence of the as-measured permittivity, $\varepsilon^*_1$, at an energy above the gap ($\hbar \omega = 0.8$ eV). In the metallic phase, $\varepsilon^*_1$ decreases with decreasing temperature, following the temperature dependence of the scattering rate $\gamma(T)$. This is characteristic of a narrowing of the Drude peak where the spectral weight is removed from the high-energy tail and transferred to the far-infrared range near the origin. The charge-gap formation below $T_{MI}$ in $N=2$ SLs leads to the reverse spectral-weight transfer from the inner-gap region to excitations across the gap, and as a consequence, to an increase in $\varepsilon^*_1$.

The consistent temperature evolution of $\varepsilon_1$ and $\varepsilon_2$ over a broad range of photon energies demonstrates the intrinsic nature of the charge-localization transition observed in SLs with $N=2$ and provides clues to its origin . The spectral-weight reduction within the gap can be quantified in terms of the effective number of charge carriers per Ni atom and extracted from a sum-rule analysis as $\Delta SW$ = $\frac{2m}{\pi e^{2} n_{\rm Ni}} \int_0^{\Omega_G} [\sigma _{1}(T\approx T_{MI},\omega)-\sigma _{1}(T<<T_{MI},\omega)] d\omega$, where $m$ is the free-electron mass and $n_{\rm Ni}$ the density of Ni atoms. The upper integration limit, $\Omega_G\approx 0.43$ eV, is a measure of the charge gap and can be identified with the equal-absorption (or isosbestic) point, where $\sigma_1(\omega)$ curves at different temperatures intersect (Fig. 2E). The charge-gap formation on this energy scale can be attributed to a charge-ordering instability, as in the case of bulk lanthanide nickelates $R\rm NiO_3$ with smaller $R$-ion radius \cite{Mazin,Munoz,Staub1,Staub2}. The spectral weight $\Delta SW \approx 0.03$ below $\Omega_G\approx 0.43$ eV determined for the $N = 2$ SLs (Fig. 2E) is of the same order as, albeit somewhat lower than, the corresponding quantity $\Delta SW =0.058$ below $\Omega_G \approx 0.3$ eV  reported at the metal-insulator transition in bulk $\rm NdNiO_3$, which is known to be due to charge order \cite{Tokura}. To highlight the analogy to the behavior in bulk nickelates, Fig. 2F shows reference measurements on single 100 nm thick films of $\rm NdNiO_3$ and $\rm LaNiO_3$, measured under the same conditions as in Figs. 2C and 2D. Since $\varepsilon^*_1$ at 0.8 eV for the single $\rm NdNiO_3$ film displays closely similar temperature dependence as found for $N = 2$ SLs, we conclude that the gap formation in the latter case also reflects charge ordering. In $\rm NdNiO_3$ the metal-insulator transition occurs as a first-order transition with a concomitant non-collinear antiferromagnetic ordering at $T_N = T_{MI}$ \cite{Munoz,Staub1,Staub2}. The thermal hysteresis in the $\varepsilon^*_1 (T)$ curve in Fig. 2F is consistent with the first-order character of the transition, with uniform and charge-ordered phases coexisting over a broad temperature range. In contrast, there is no discernible hysteresis observed in $\varepsilon^*_1 (T)$ of $N=2$ SLs (Figs. 2C and 2D), which suggests a second-order transition. Continuing the analogy with the bulk nickelate series, one would then expect another second-order transition due to the onset of antiferromagnetic ordering at $T_N < T_{MI}$ in the $N=2$ SLs, as in $R$$\rm NiO_3$ with small $R$ (Lu through Sm).

In order to test this hypothesis, we carried out low-energy muon spin rotation (LE-$\mu$SR) measurements using the $\mu$E4 beamline at the Paul Scherrer Institute \cite{Thomas,SOM}, where positive muons with extremely reduced velocity can be implanted into specimens and brought to rest between the substrate and the LaAlO$_3$ capping layer. Since the muons decay into positrons preferentially along the spin direction, they act as highly sensitive local magnetic probes. Figure 3A shows muon decay asymmetry data from a SL with $N = 2$ at selected temperatures with no external field. At $T >$ 50 K, the asymmetry is described by a Gaussian with relatively slow relaxation, $\sigma$, given by $A(0)\times\exp(-\sigma^2t^2/2)$ (solid lines in Fig. 3A), typical of dipolar magnetic fields generated by nuclear moments of La and Al. As the temperature decreases, there is a gradual increase in $\sigma$ from 0.17 $\mu $s$^{-1}$ at 250 K to 0.27 $\mu $s$^{-1}$ at 20 K. Below 50 K the asymmetry drops sharply, and the $\mu$SR spectra can be fitted well by introducing an additional exponential relaxation $\exp(-\Lambda t)$. The fast depolarization rate $\Lambda$ reaches a value of $\approx 17 \mu $s$^{-1}$ at 5 K, implying a resulting Lorentzian distribution of local fields with half-width-at-half-maximum $\Delta B = 0.75\Lambda/\gamma_{\mu} \approx 150$ G, where $\gamma_{\mu } = 2\pi\times 13.55$ MHz/kG is the muon's gyromagnetic ratio \cite{SOM}. The fast increase in $\Lambda $ with decreasing temperature below 50 K is similar to the behavior in bulk $\rm NdNiO_3$  \cite{Munoz2} and $\rm (Y,Lu)NiO_3$ \cite{Munoz3} below $T_N$, caused by static internal fields from ordered Ni magnetic moments. The wide field distribution $\Delta B$ and the absence of a unique muon precession frequency reflects the SL structure with several inequivalent muon stopping sites in the alternating magnetic (LaNiO$_3$) and nonmagnetic (LaAlO$_3$) layers, probably compounded by a complex non-collinear spin structure as in the bulk nickelates \cite{Munoz2,Munoz3}. 

We used 100 G transverse field (TF) measurements to determine the fraction of muons, $\rm f_m$, experiencing static local magnetic fields $B_{\rm loc}>B_{\rm TF}$ (i.e. showing no detectable precession with $\omega = \gamma_{\mu}B_{\rm TF}$) \cite{SOM}. Figure 3C indicates that the $N = 2$ SL shows a transition from an entirely paramagnetic muon environment ($\rm f_m = 0$) to a nearly full volume of static internal fields, with a sharp onset at $T_N \sim $ 50 K. The magnetic state at 5 K is robust against externally applied transverse fields up to 3 kG (not shown), that is the limit of time resolution of our setup. The continuous temperature dependence of $\rm f_m$ (Fig. 4) and the absence of thermal hysteresis indicate that the magnetic transition for the $N=2$ SLs is second-order. At the same time, Figs. 3B and 3D show that SLs with thicker $\rm LaNiO_3$ layers remain paramagnetic down to the lowest temperatures, as in bulk $\rm LaNiO_3$. An additional slow exponential relaxation with $\Lambda = 0.9\  \mu $s$^{-1}$ is seen only at $T= 5$ K (black symbols and curve in Fig. 3B). This results in a small increase in relaxation rate, but no loss in asymmetry of the TF $\mu $SR signal (Fig. 3D). The effect is likely due to weak dynamical spin correlations that are quenched already in a field of 100 G, in clear contrast to the long-range static magnetic order observed in the $N=2$ samples.

As a local probe, $\mu$SR does not allow definite conclusions about the magnetic ordering pattern in the $N=2$ SLs. However, we can rule out ferromagnetism based on an estimate of the ordered moment, $\mu_{Ni}$, on the Ni sites from the distribution of local fields experienced by the muons. The highest local field at the shortest $\mu$-Ni distance, $c/4$ (or $c^{\ast}/2$ in pseudo-cubic notation) $\approx $ 1.92 \AA \cite{Munoz2}, is 4-5 times $\Delta B$, which corresponds to $\mu_{Ni}\gtrsim 0.5\mu_B$. If these moments were co-aligned in the ordered state, the corresponding total moment $M=\mu_{Ni}n_{\rm Ni}V_{SL} \gtrsim 7.7\times 10^{-4}$ emu would have been readily detected in magnetization measurements. The absence of such an effect is confirmed in magnetometric measurements with sensitivity $\sim 10^{-7}$ emu.

Figure 4 summarizes the phase behavior of the SLs with $N=2$, which undergo a sequence of two sharp, collective electronic phase transitions upon cooling. We have provided strong evidence that the two transitions correspond to the onset of charge and spin order. By showing that the $N=4$ counterparts remain uniformly metallic and paramagnetic at all temperatures, we have demonstrated full dimensionality control of these collective instabilities. The higher propensity towards charge and spin order in the two-dimensional systems probably reflects enhanced nesting of the LaNiO$_3$ Fermi surface. The phase behavior is qualitatively similar to the one observed in bulk $R$NiO$_3$ with small radius of the $R$ anions, which results from bandwidth narrowing due to rotation of NiO$_6$ octahedra, but the transition temperatures and the order parameters are substantially lower, probably because of the reduced dimensionality. Since the transitions occur in the $N=2$ SLs irrespective of whether the substrate-induced strain is compressive (Fig. 1B) or tensile (Fig. 1C), structural parameters such as rotation and elongation of the NiO$_6$ octahedra can be ruled out as primary driving forces. We note, however, that the infrared conductivity is higher (Figs. 2A and 2B) and the transition temperatures are lower (Fig. 4) in the $N = 2$ SL grown under tensile strain. The more metallic response of these SLs, compared to those grown under compressive strain, may reflect a widening of the Ni 3d bandwidth and/or an enhanced occupation of the Ni $d_{x^2-y^2}$ orbital polarized parallel to the LaNiO$_3$ layers. A small orbital polarization was indeed detected by soft x-ray reflectometry in our superlattices \cite{benckiser}. This indicates further opportunities for orbital control of the collective phase behavior of the nickelates, which may enable experimental tests of theories predicting high-temperature superconductivity \cite{Giniyat,Hansmann} or multiferroicity \cite{Brink} in these systems.

\newpage

\begin{figure*}
\includegraphics[width=12cm]{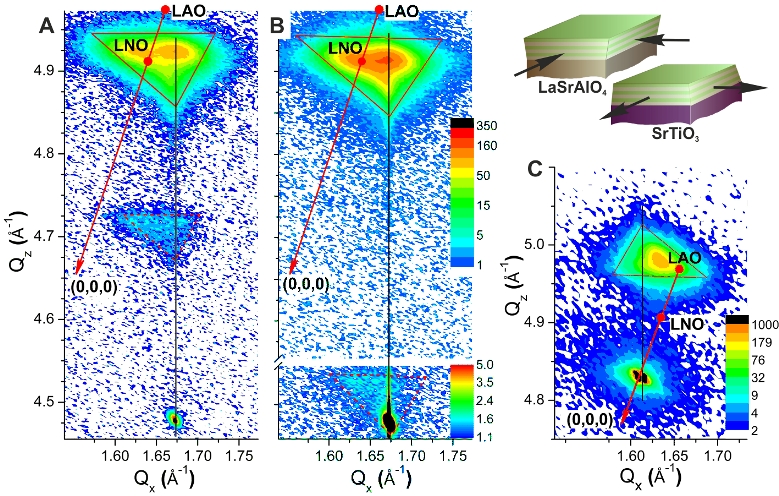}
\caption{Reciprocal-space maps of 100-nm-thick \SLN \ superlattices grown under compressive strain on \LSAO \ with (\textsf{\textbf{A}}) $N = 4$ and (\textsf{\textbf{B}}) $N = 2$ and (\textsf{\textbf{C}}) under tensile strain on \STO \ with $N = 2$. The black vertical lines indicate the in-plane ($\rm Q_x$) position of the $\rm LaSrAlO_4$ ($109$) and $\rm SrTiO_3$ ($103$) reflections. The strain state of the perovskite epilayers is identified by the intensity distribution in the vicinity of the ($103$) layer Bragg peak and its superlattice satellite, which are delineated by solid- and dashed-line triangles, respectively. The reciprocal spacings of $103$ strain-free pseudo-cubic \LNO \ and \LAO \ are indicated by the red circles. The red arrows point towards the origin. }
\protect\label{Fig1}
\end{figure*}

\begin{figure*}
\includegraphics[width=12cm]{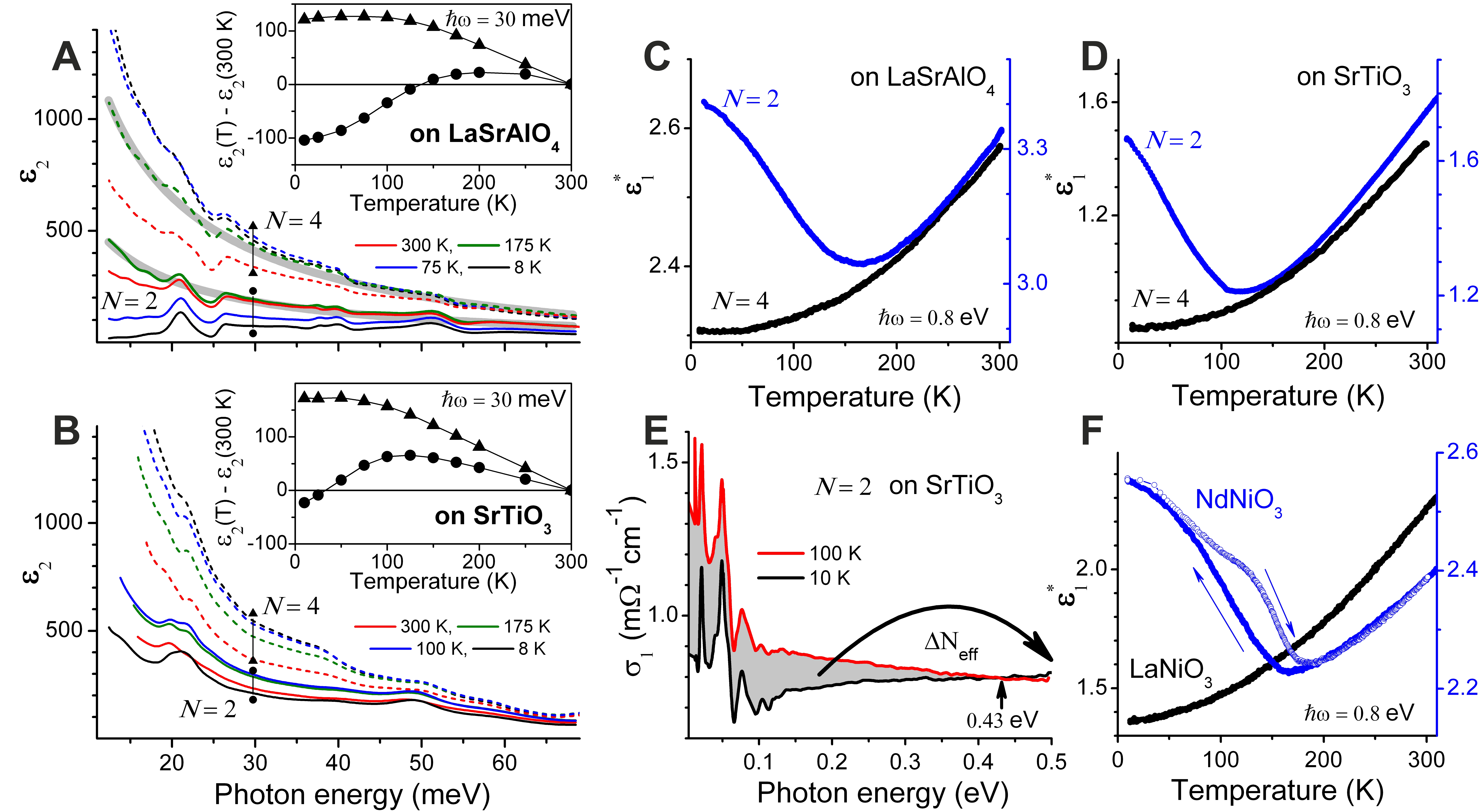}
\caption{(\textsf{\textbf{A,B}}) $\varepsilon_2(\omega)$ spectra of the $N = 2$ (solid lines, circles) and $N = 4$ (dashed lines, triangles) SLs on (\textsf{\textbf{A}}) $\rm LaSrAlO_4$ and (\textsf{\textbf{B}}) $\rm SrTiO_3$ substrates measured at representative temperatures. The shaded lines in (\textsf{\textbf{A}}) represent the Drude model fit to $\varepsilon_2(\omega)$ at 175 K for the $N = 2$ ($N = 4$) SL with $\omega_{pl}=$ 1.05 eV (1.10 eV) and $\gamma =$ 200 meV (90 meV). The insets provide the corresponding temperature dependencies of $\varepsilon_2$ at a photon energy of $\hbar \omega = 30$ meV for the $N = 2$ (circles) and $N = 4$ (triangles) SLs. (\textsf{\textbf{C,D}}) Temperature dependence of the as-measured pseudo-dielectric permittivity $\varepsilon^*_1$ at $\hbar \omega = 0.8$ eV in the $N = 2$ (blue) and $N = 4$ (black) SLs on (\textsf{\textbf{C}}) $\rm LaSrAlO_4$ and (\textsf{\textbf{D}}) $\rm SrTiO_3$. (\textsf{\textbf{E}}) The difference between the optical conductivity spectra $\sigma_1($100 K, $\omega)$ and $\sigma_1($10 K, $\omega)$ (shaded area) quantifies the reduction of the effective charge density, $\Delta SW \approx$ 0.03 per Ni atom, within the gap energy range below 0.43 eV at the charge ordering transition in the $N = 2$ superlattice on $\rm SrTiO_3$. (\textsf{\textbf{F}}) Temperature dependence of $\varepsilon^*_1$ at 0.8 eV of reference 100 nm films of $\rm LaNiO_3$ (black) and $\rm NdNiO_3$ (blue).
}
\protect\label{Fig2}
\end{figure*}

\begin{figure*}
\includegraphics[width=8.9cm]{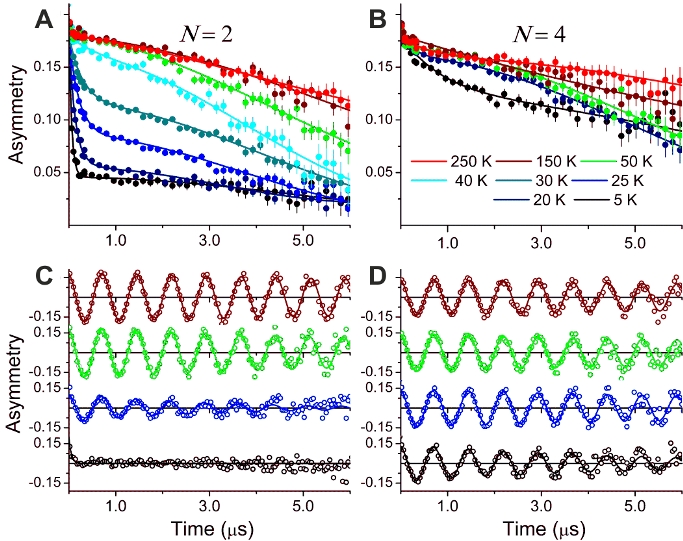}
\caption{The time evolution of
the zero-field muon spin polarization at various temperatures for the (\textsf{\textbf{A}}) $N = 2$ and (\textsf{\textbf{B}}) $N = 4$ superlattice on $\rm LaSrAlO_4$. (\textsf{\textbf{C}}) and (\textsf{\textbf{D}}) Muon spin relaxation spectra in a weak transverse magnetic field of 100 G in the superlattices as in (\textsf{\textbf{A}}) and (\textsf{\textbf{B}}), respectively. (\textsf{\textbf{A}}) to (\textsf{\textbf{D}}) use the color coding in legend (\textsf{\textbf{B}}).
}
\protect\label{Fig3}
\end{figure*}

\begin{figure*}
\includegraphics[width=5.8cm]{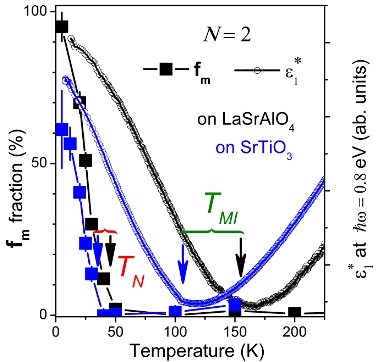}
\caption{Temperature dependencies of the fraction of muons experiencing static local magnetic fields, $\rm f_m$, and the normalized permittivity $\varepsilon^*_1$ at 0.8 eV in the $N = 2$ superlattices. The black and blue arrows mark the magnetic ($T_N$) and metal-insulator ($T_{MI}$) transition temperatures for the superlattices on $\rm LaSrAlO_4$ and $\rm SrTiO_3$, respectively.
}
\protect\label{Fig4}
\end{figure*}


\begin{thebibliography}{50}
\bibitem{SOM} Materials and methods are available as supporting materials on Science Online.
\end{thebibliography}
\end{document}


\baselineskip22pt
\maketitle

\renewcommand\thefigure{S\arabic{figure}}
\renewcommand{\cite}[1]{(S\citenum{#1})}
\renewcommand{\bibnumfmt}[1]{[S#1]}

\section*{Methods and materials}

\paragraph*{Sample preparation and characterization\\}
High-quality superlattices (SLs) composed of $N$ u.c. thick consecutive layers of $\rm LaNiO_3$ and $\rm LaAlO_3$ were grown by pulsed-laser deposition from stoichiometric targets using a KrF excimer laser with 2\,Hz pulse rate and 1.6\,J/cm$^2$ energy density. Both compounds were deposited in 0.5 mbar oxygen atmosphere at 730$^{\circ}$C and subsequently annealed in 1 bar oxygen atmosphere at 690$^{\circ}$C for 30 min. We have grown SLs on two kinds of single-crystalline
substrates: $\rm SrTiO_3$, which induces tensile strain in the overlayer, and $\rm LaSrAlO_4$, which induces compressive strain (see Table SI).
All substrates were 10\,mm$\times$10\,mm$\times$0.5\,mm or 5\,mm$\times$5\,mm$\times$0.5\,mm (001)-oriented plates with a miscut angle $<$ 0.1$^{\circ}$.
\begin{table}[t]
\centering
\renewcommand{\thetable}{SI}
\caption{Average lattice constants of 100 nm thick $N = 2$ SLs grown
on (001)-oriented $\rm SrTiO_3$ and $\rm LaSrAlO_4$ substrates (determined from the main (103) layer Bragg peak positions in Figs. 1B and 1C) in comparison with the lattice constants of strain-free pseudo-cubic $\rm LaNiO_3$ and $\rm LaAlO_3$ and the same substrates.}
\begin{tabular}{ccccccc}
\hline\hline
 & SL on LaSrAlO$_4$ & SL on SrTiO$_3$ & $\rm LaNiO_3$& $\rm LaAlO_3$& $\rm LaSrAlO_4$ & $\rm SrTiO_3$\\
\hline
$a,b$ (\AA)& 3.769& 3.845& 3.837&3.789&3.756&3.905 \\
$c$ (\AA)&3.853&3.790& 3.837& 3.789&12.636&3.905\\
\hline\hline
\end{tabular}%
\end{table}
We chose to work on 100 nm thick SLs in order to enhance the dielectric response and to confine the muon stopping distribution within the SL. The chosen thickness range also allows us to avoid complications arising from initial growth of TMO layers on a substrate \cite{Jak}.
The growth rates for the individual layers were controlled
by counting laser pulses in combination with feedback from high-resolution x-ray diffraction measurements. The crystallinity, superlattice structure, and sharpness of the interfaces (with roughness $<$ 1 u.c.) were verified by momentum-dependent x-ray reflectivity and high-resolution hard x-ray diffraction
\begin{figure}[h]
\includegraphics*[width=\columnwidth]{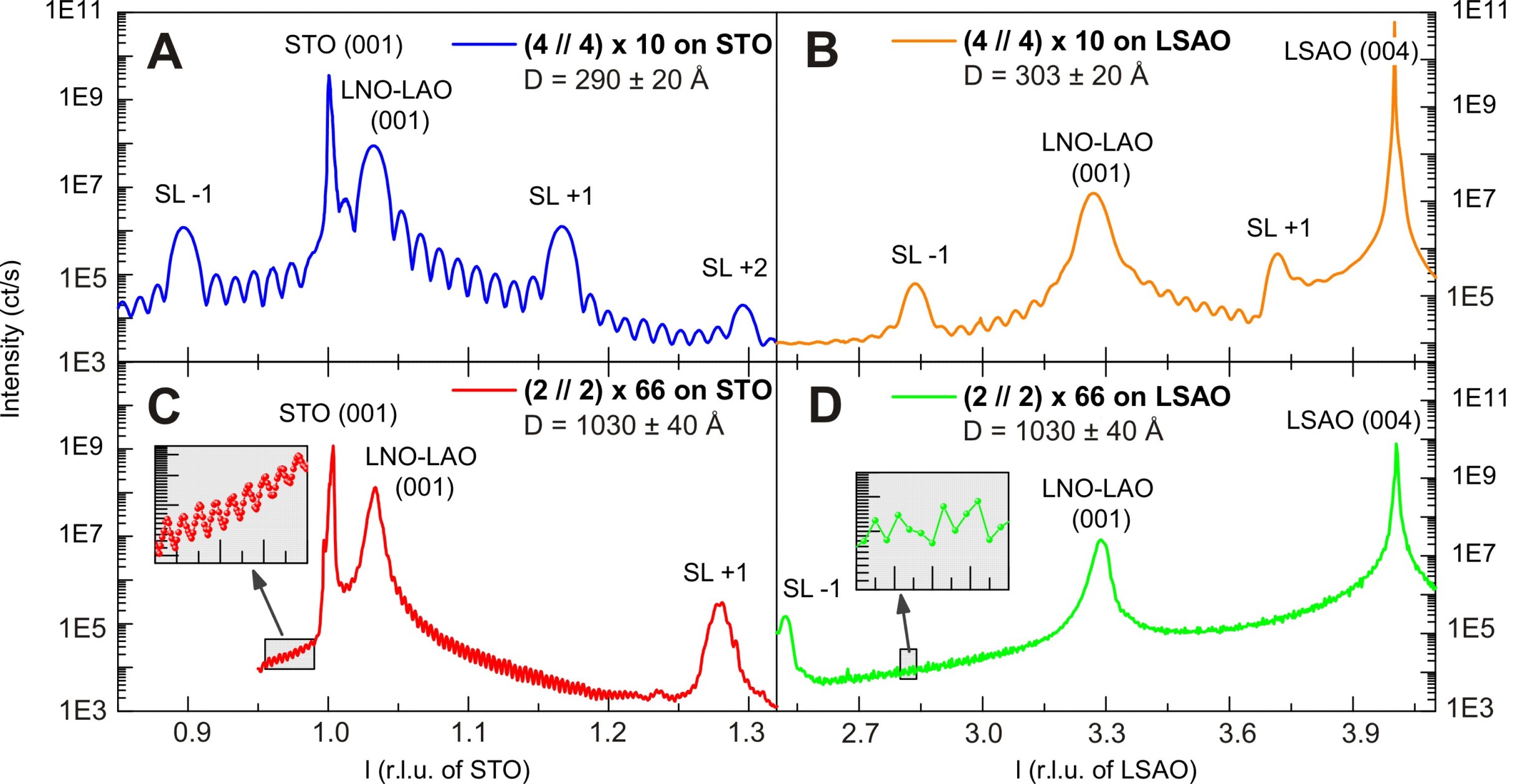}
\caption{
High-resolution x-ray diffraction measured with 10\,keV synchrotron radiation at the MPI-MF beamline of the ANKA facility at the Karlsruhe Institute of Technology for the (A, B) 30 nm ($N=4$) and (C, D) 100 nm ($N=2$) thick superlattices on (A,C) $\rm SrTiO_3$ and (B, D) $\rm LaSrAlO_4$. The thickness of the SL in (D) is determined from the hard x-ray reflectivity measurements in Fig. 2D.  The 100 nm thick samples were used for low-energy muon spin rotation and ellipsometry experiments.
\label{XRD}}
\end{figure}
\begin{figure}[htb]
\includegraphics*[width=1.00\columnwidth]{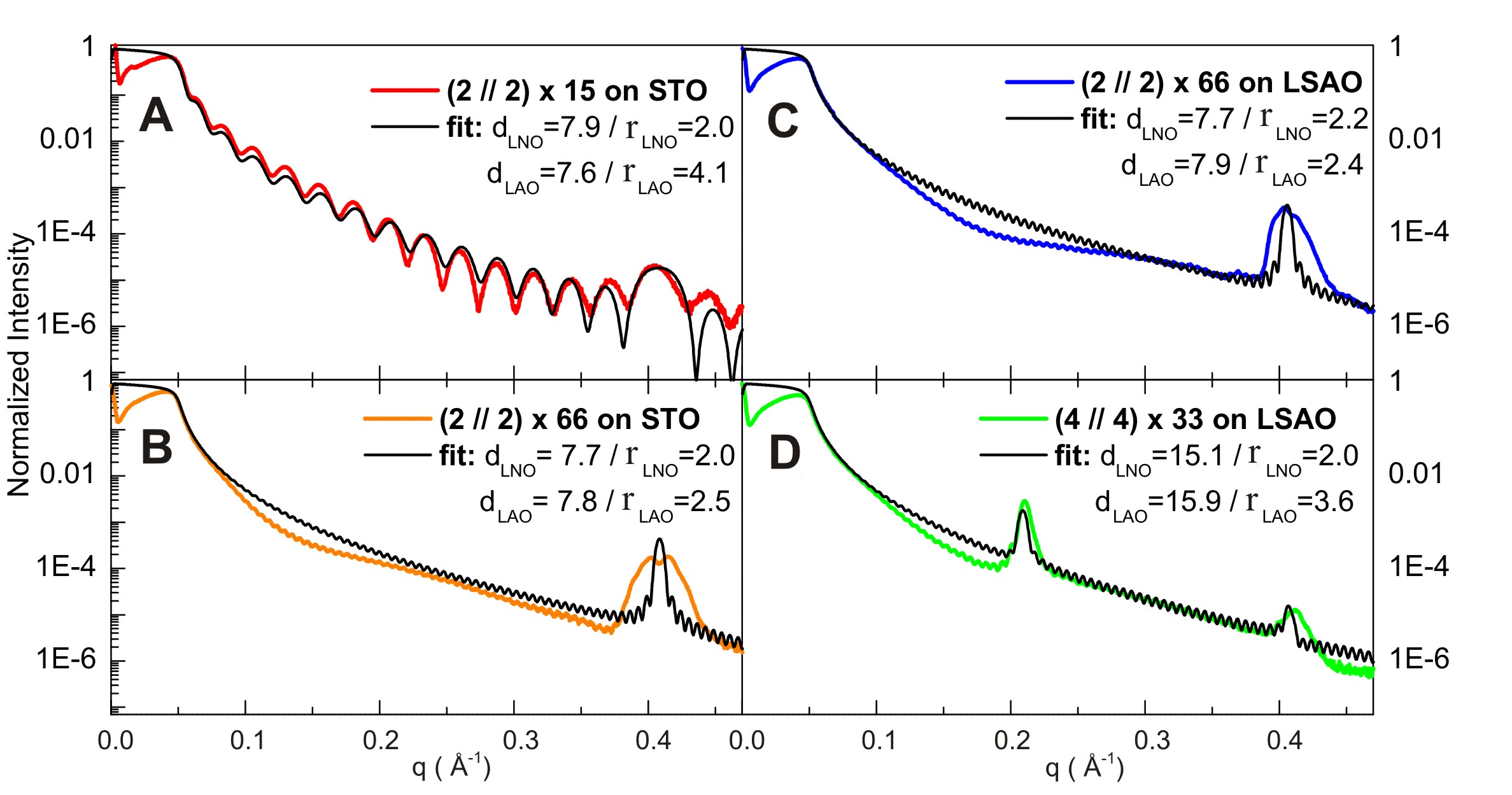}
\caption{
Hard x-ray reflectivity measured with Cu $K\alpha $ radiation and fits for the (A) 23 nm and (B) 100 nm thick $N=2$ superlattices on $\rm SrTiO_3$ and 100 nm thick (C) $N=2$ and (D) $N=4$ superlattices on $\rm LaSrAlO_4$. The samples in (C-D) were used for low-energy muon spin rotation and ellipsometry experiments. \label{Refl}}
\end{figure}
\begin{SCfigure}[][htb]
\includegraphics*[width=0.56\columnwidth]{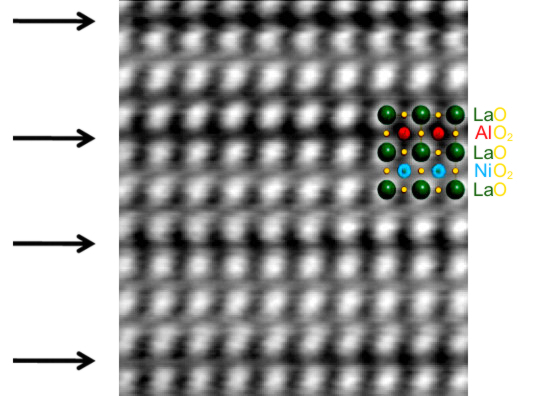}
\caption{High-angle annular dark field image of the $\rm LaNiO_3\ ( 2\ u.c.)| LaAlO_3\ (1\ u.c.)$ superlattice. Subsequent dark (marked by arrows) and bright layers show the chemical variation of the layer system.
\label{TEM1}}
\end{SCfigure}
scans which revealed, besides the perovskite Bragg reflections, satellite peaks due to the long-range multilayer superstructure and Kiessig fringes caused by total-thickness interference.

Representative scans along the specular truncation rod are shown in Fig.~\ref{XRD} for samples grown on the different substrates with different individual layer thicknesses $N$ (u.c.), and total thicknesses $D$ (\AA). Symmetrically around the (001) layer Bragg peak one can see superlattice satellites and $M-2$ thickness fringe maxima, where $M$ is the number of bilayer repetitions. The position of the satellites corresponds to a \SLN \ bilayer thickness of 30\,$\pm$\,1\,\AA\,  and 15.5\,$\pm$\,0.5\,\AA\, for the $N=4$ (Figs. S1A and S1B) and $N=2$ (Figs. 1C  and 1D), respectively, that is in a good agreement with the $2 N c$ value, where $c$ is the average epilayer lattice constant in Table SI. Accordingly, the Kiessig fringes in Figs. S1 and S2 correspond to the total thickness $M\times 2 N c$. The thickness fringes for the 100 nm thick $N = 2$ SL on $\rm LaSAlO_4$ are damped at higher $l$ values in Fig. S1D, but well resolved in Fig. S2D. The x-ray reflectivity shown in Fig. S2 was also used to characterize the superlattice structure and sharpness of the interfaces. From fits to the reflectivity, using the Parratt algorithm and tabulated values for the optical constants
\renewcommand{\cite}[1]{S\citenum{#1}} (\cite{ReMagX},\cite{Parratt1954},
\cite{Chantler})\renewcommand{\cite}[1]{(S\citenum{#1})}, we obtained the thickness ($d_{LAO},\ d_{LNO}$) and roughness ($\sigma_{LAO},\ \sigma_{LNO}$) of the individual layers. Using a minimal set of fitting parameters (assuming $M$ identical $\rm LaNiO_3$ and $\rm LaAlO_3$ layers), we show in Fig. S2 a good description of the data. The roughness parameters are all around 1\,u.c or less and represent values averaged over a large area of $\sim$(10$\times$1)\,mm which corresponds to the x-ray spot size and inevitably contains planar defects such as stacking faults. This indicates the presence of atomically flat and abrupt interfaces. Some of the samples were also checked by high-resolution transmission electron microscopy (TEM), providing a local picture of the atomic stacking sequence. In Fig. S3 a high-angle annular dark-field image of a $\rm LaNiO_3\ ( 2\ u.c.)| LaAlO_3\ (1\ u.c.)$ SL is shown.
In this imaging mode, also known as $Z$-contrast, the contrast is proportional to $Z^n$, where $Z$ is the atomic number and $n$ is about 1.7. Subsequent dark (marked by arrows) and bright layers show a chemical variation of the layer system. In this example a sequence of two $\rm LaNiO_3$ layers and one $\rm LaAlO_3$ layer is visible which shows that even single layers can be deposited without distinct intermixing.
The superior quality of our samples is also supported by resonant reflectivity measurements performed on a sample grown under the same conditions. The analysis of those data allowed some of us to determine the atomic-layer resolved orbital polarization in these superlattices \cite{Benckiser}.

\paragraph*{Substrate-induced strain and relaxation effects\\}

In general, the physical properties of thin films are strongly influenced by substrate-induced strain and relaxation effects. It has thus far proven difficult to separate the influence of the dimensionality from that of other parameters such as the strain-induced local structural distortions and interfacial defects. In order to discriminate between these effects we chose to work on SLs grown on both $\rm SrTiO_3$, which induces tensile strain in the overlayer, and $\rm LaSrAlO_4$, which induces compressive strain. Our comprehensive reciprocal-space mapping (RSM) measurements \cite{Frano} supplemented by high-resolution TEM micrographs verified that strain and relaxation effects are strongly affected by inversion of the type of substrate-induced strain, but remain essentially unchanged by varying the individual layer thicknesses. In our study, we show that, on the contrary, the transport and magnetic properties of the SLs are almost unaffected by inversion of the type of substrate-induced strain, but qualitatively transformed by varying the number of consecutive unit cells within the LaNiO$_3$ layers. Since the metal-insulator and spin-ordering transitions occur in the $N = 2$ SLs irrespective of whether the substrate-induced strain is compressive or tensile, strain-induced local structural distortions and interfacial defects are ruled out as primary driving forces.
\begin{figure}[h]
\includegraphics*[width=1.0\columnwidth]{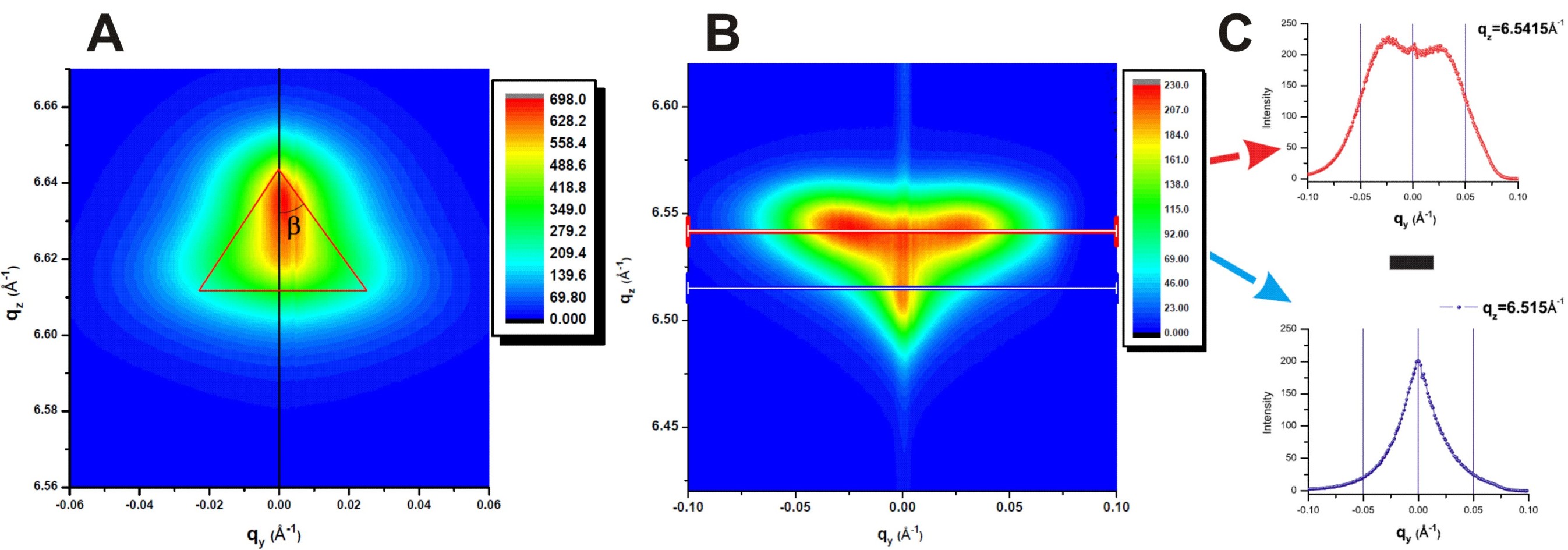}
\caption{Reciprocal-space maps in the vicinity of the symmetric (004) peak of the 100 nm thick $N=2$ superlattices on (A) $\rm SrTiO_3$ and (B) $\rm LaSrAlO_4$ substrates. The relaxation triangle is highlighted with a red line in (A). The angle $\beta \approx 2^{\circ}$ quantifies the amount
of gradual relaxation the SL has. (C) Horizontal cuts along the indicated in (B) $q_z$ values. The separation of the two twin peaks reveal the formation of twin domains.
\label{RSM}}
\end{figure}
\begin{figure}[htb]
\includegraphics*[width=1.0\columnwidth]{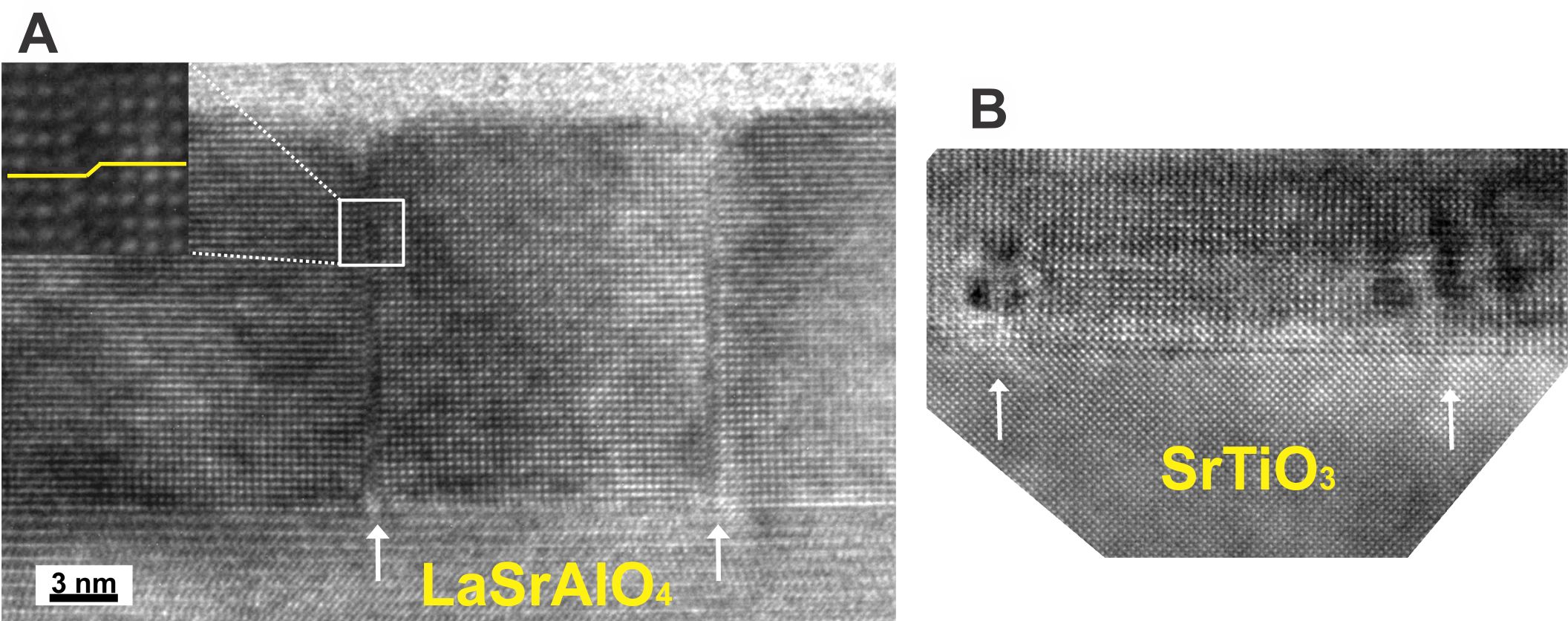}
\caption{High-resolution TEM micrographs of $\rm LaAlO_3| LaNiO_3$ SLs on (A) $\rm LaSrAlO_4$ and (B) $\rm SrTiO_3$ substrates. Defects are marked by arrows. The inset in (A) shows a magnified area close to a planar defect. \label{TEM2}}
\end{figure}

Figure 1 of the main text shows contour maps of the diffracted X-ray intensity distribution in the vicinity of the $103$ perovskite Bragg peak for three representative samples: $N = 4$ and $N = 2$ SLs grown on \LSAO, and an $N = 2$ SL on \STO.
The analysis of the averaged in-plane and out-of-plane lattice constants (Table SI) indicates that compressive strain reduces the in-plane lattice parameter by $\Delta a/a \approx 1.8\ \%$ relative to the bulk $\rm LaNiO_3$ lattice, whereas tensile strain results in a reduction of the out-of-plane lattice constant by $\Delta c/c \approx 1.2\ \%$. These two types of local distortions in the perovskite structure are accommodated by rotations of the $\rm NiO_6$ octahedra about different Cartesian axes \cite{May}, which, in turn, exert an inequivalent influence on the $\rm LaNiO_3$ electronic structure. A distribution of the diffracted intensity near the epilayer reflection for SLs grown on $\rm LaSrAlO_4$ has a characteristic triangular shape, with dispersion along the in-plane ($Q_x$) direction towards the $103$ Bragg reflection of strain-free bulk $\rm LaNiO_3$. This is in contrast to the tensile-strained SLs grown on $\rm SrTiO_3$, where the strain relaxation is characterized by nearly elliptical contour lines close to the $103$ Bragg reflection of cubic $\rm LaAlO_3$. The tensile strain of $\rm SrTiO_3$ is $(67\pm3)\%$ relaxed, and by comparing with SLs of less total thickness, we identified a faint gradient-profile-relaxation effect as function of overlayer thickness, similar to the behavior observed in semiconductor heterostructures \renewcommand{\cite}[1]{S\citenum{#1}} (\cite{Heinke1}, \cite{Heinke2}).\renewcommand{\cite}[1]{(S\citenum{#1})} In addition to Fig. 1 of the main text, Fig. S4 shows RSMs of the symmetric 004 perovskite Bragg peak measured with synchrotron radiation at the MPI-MF beamline of the ANKA facility at the Karlsruhe Institute of Technology. The diffracted x-ray intensity distribution for the $N=2$ SL on $\rm SrTiO_3$ (Fig. S4A) exhibits the triangular shape described in Refs.\renewcommand{\cite}[1]{S\citenum{#1}} \cite{Heinke1} and \cite{Heinke2}.\renewcommand{\cite}[1]{(S\citenum{#1})}
The effect of triangular relaxation was not observed in the
thinner SL (not shown). Because the lattice constants of the thin SL are almost equal to the ones of the thick sample, the SLs grown on $\rm SrTiO_3$ seem to relax abruptly at the beginning of the growth. Further, from the peak shape evolution, the SL gradually relaxes the tensile strain.  The subtle thickness evolution of the layer's relaxation indicates that it is the substrate surface where abrupt strain-adapting mechanisms take place.  The effect of tensile strain on TMO heterostructures may produce oxygen vacancies \cite{Conchon},
which give rise to a different valence state of the Ni ion at the substrate interface \cite{Jak}.  Figures S4B and S4C show  that the distribution of the diffracted intensity near the epilayer reflection for SLs grown on $\rm LaSrAlO_4$ has a double-peak splitting along the in-plane ($Q_x$) direction. This intensity pattern (only seen in thicker SLs grown on $\rm LaSrAlO_4$) suggests the formation of twinning domains, as described in Refs.\renewcommand{\cite}[1]{S\citenum{#1}} \cite{Gebhardt1}, \cite{Gebhardt2}.\renewcommand{\cite}[1]{(S\citenum{#1})}
The two different relaxation mechanisms in the perovskite structure are confirmed by TEM measurements performed on samples grown under the same conditions as in our study. Figure S5 shows high-resolution TEM micrographs (recorded by a JEOL JEM4000FX microscope) of the $\rm LaNiO_3 - LaAlO_3$ layer systems.  In the case of the $\rm LaSrAlO_4$ substrate (Fig. S5A) planar defects are visible (marked by arrows)  which are oriented perpendicular to the substrate plane and extend through the entire SL. As shown in the magnified inset image, the stacking sequence changes at these faults (yellow broken line).  The size of the defect-free blocks varies between 15 and 50 nm. The microstructure of the layer system on the $\rm SrTiO_3$ substrate (Fig. S5B) only very occasionally shows planar defects. Instead, localized defects are found close to the substrate (marked by arrows). These defects can be associated with the creation of oxygen vacancies and changes in the oxygen coordination of Ni ions at the substrate interface. Recent photon energy-dependent hard x-ray photoelectron spectroscopy measurements on some of our samples have confirmed that the initial growth on the $\rm SrTiO_3$ surface leads to the $\rm Ni^{2+}$ valence state \cite{Claessen}.
The oxygen vacancy formation energy gradually decreases with increasing the in-plane perovskite lattice spacing \cite{Kotomin}, which can explain the marked difference in the oxygen vacancy concentration in thin films grown under tensile or compressive strain \cite{Conchon}. Nevertheless, in our study, the temperature-induced phase transitions occur in the $N = 2$ (but not in $N = 4$) SLs irrespective of whether the substrate-induced strain is compressive or tensile, which clearly distinguishes these transitions from those in highly oxygen deficient $\rm LaNiO_{3-\delta}$ ($\delta \geq0.25$)
\renewcommand{\cite}[1]{S\citenum{#1}} (\cite{Sanchez}, \cite{Kawai}).\renewcommand{\cite}[1]{(S\citenum{#1})} Moreover, the reduced insulating phases require more than 1/3 of divalent $\rm Ni^{2+}$ in square planar (vs. perovskite octahedral) sites. Based on the detailed characterization of our samples by means of XRD, XAS, RSM, HAXPES, and TEM we can definitively rule out such a scenario.

In conclusion, our analysis confirms the excellent quality of the synthesized SLs, which exhibit abrupt interfaces and excellent crystallinity. Defect-free, atomically precise 15-50 nm blocks are separated by $\sim $ 1 u.c. stacking faults. These planar defects are inevitably caused by strain relaxation effects, and can block the current flow through the atomically thin layers. We have therefore used advanced local probes, such as spectroscopic ellipsometry and low-energy muons, to study the intrinsic electronic transport and magnetic properties of the heterostructures.

\paragraph*{Spectroscopic ellipsometry measurements and data analysis\\}

\begin{figure}[htb]
\includegraphics*[width=1.0\columnwidth]{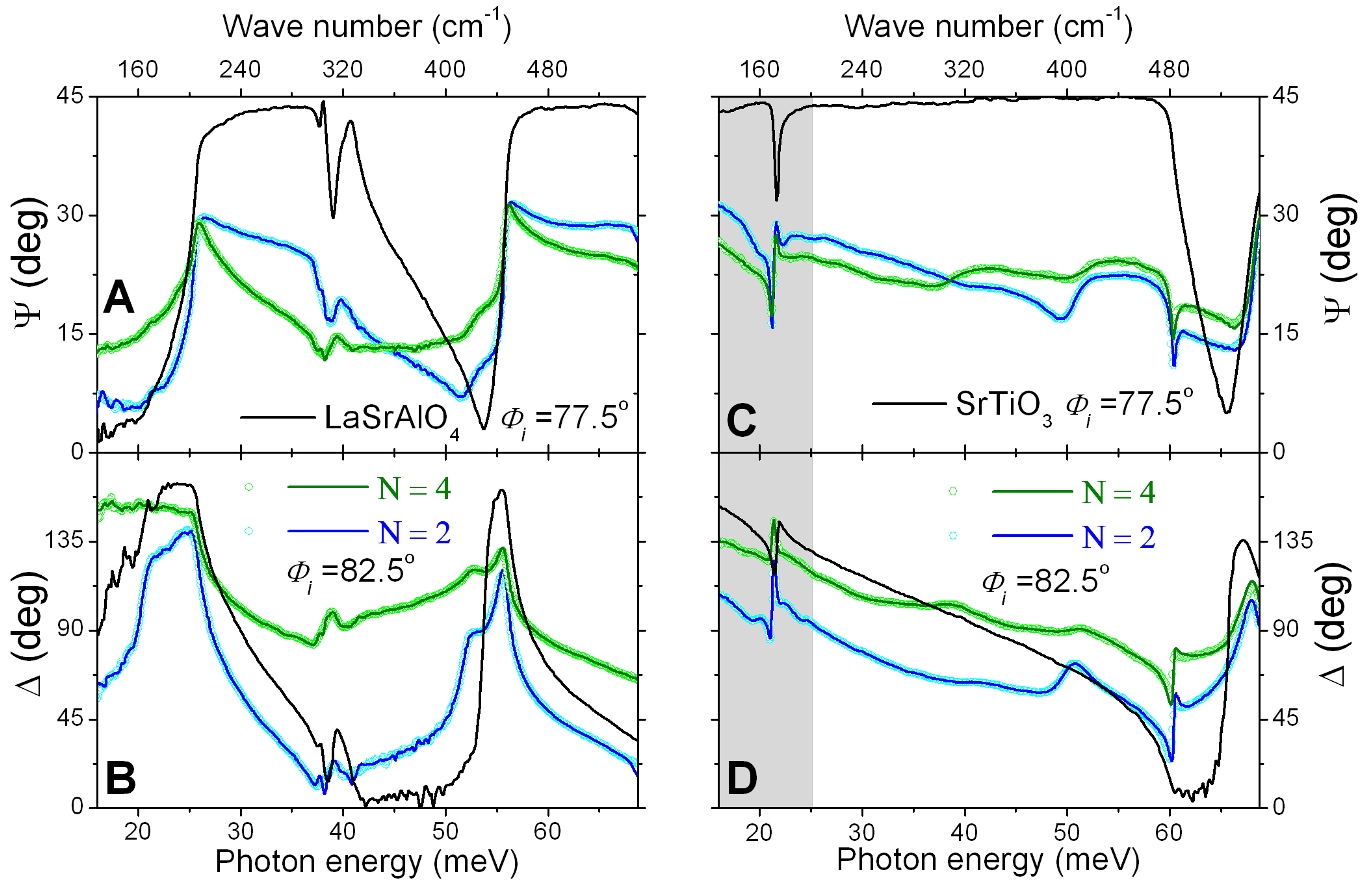}
\caption{Experimental (open circles) and best-fit calculated (solid lines)
ellipsometry spectra of the N = 2 and N = 4 SLs on (A,B) $\rm LaSrAlO_4$ at $T=175$ K and (C,D) $\rm SrTiO_3$ at $T=100$ K. The angle of incidence of the polarized light was $\Phi_i=82.5^{\circ}$.  Ellypsometry
spectra of the bare substrates measured at $\Phi_i=77.5^{\circ}$ are shown for comparison (black solid lines). The gray  shaded area in (C,D) indicates the region where the data analysis is affected by dielectric microwave dispersion of the ferroelectric soft mode of $\rm SrTiO_3$.
\label{ElliS1}}
\end{figure}
\begin{figure}[htb]
\includegraphics*[width=1.0\columnwidth]{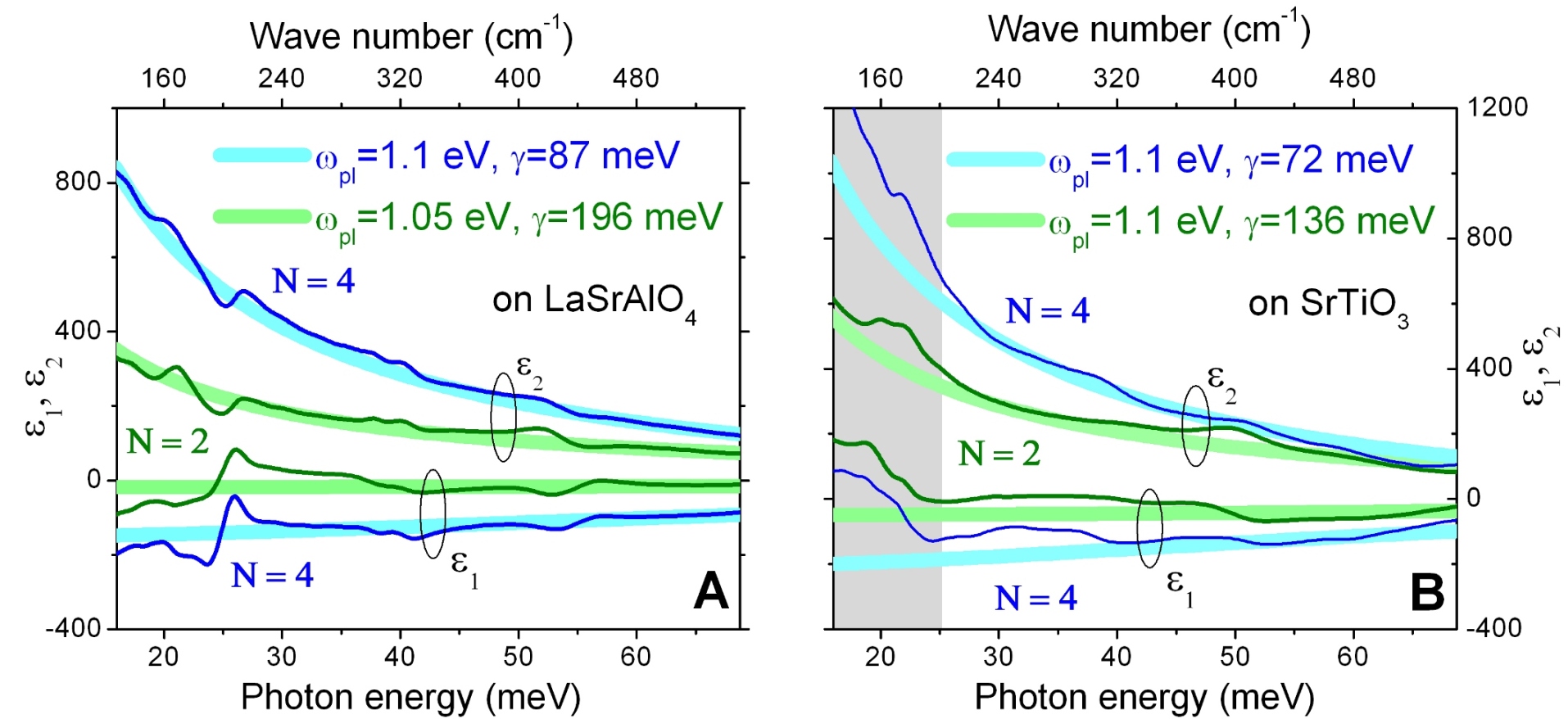}
\caption{Best-fit model functions $\varepsilon_1 (\omega)$ and $\varepsilon_2(\omega)$
for the N = 2 and N = 4 SLs on (A) $\rm LaSrAlO_4$ at $T=175$ K and (B) $\rm SrTiO_3$ at $T=100$ K, as obtained by inversion of the ellipsometric parameters in Fig. S6. The shaded lines represent the Drude model simultaneous fit to both $\varepsilon_1(\omega)$ and $\varepsilon_2(\omega )$ with parameters $\omega_{pl}$ and $\gamma $
described in the legends. The gray shaded area in (B) indicates the region where the model fitting curves deviate significantly from the data.
\label{ElliS2}}
\end{figure}
\begin{figure}[htb]
\includegraphics*[width=1.0\columnwidth]{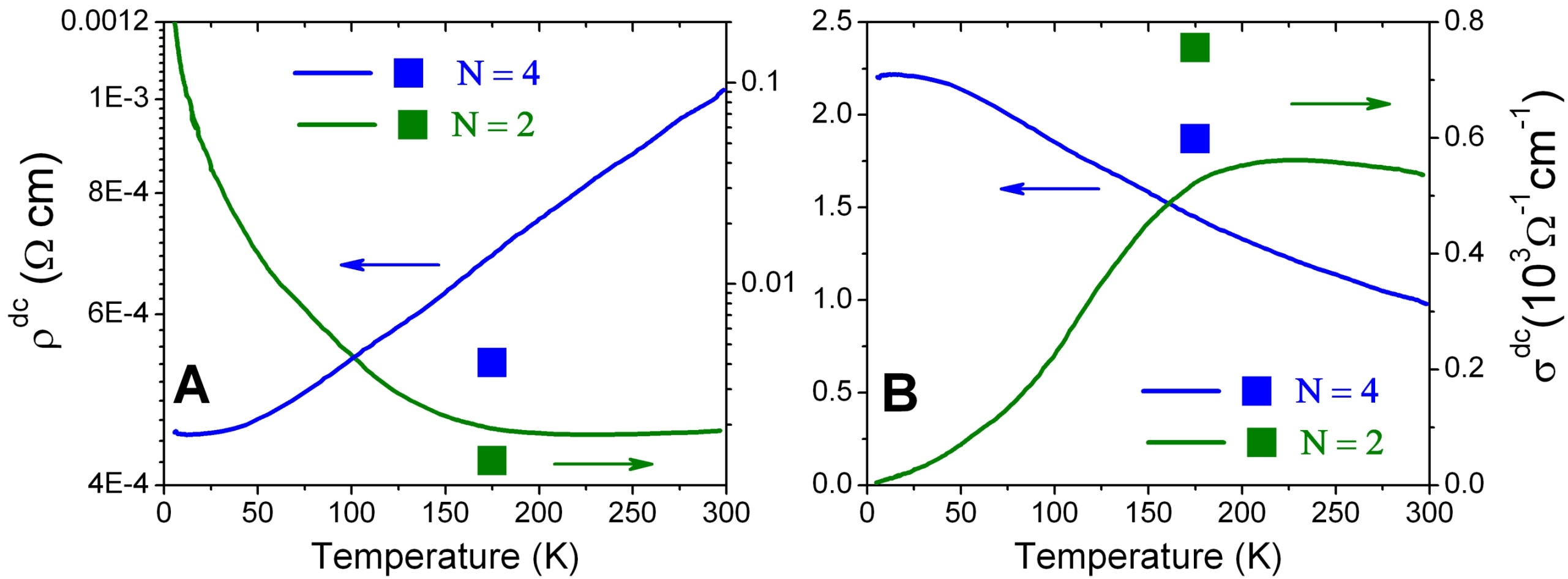}
\caption{Temperature dependence of the $dc$ (A) resistivity and (B) conductivity of the N = 2 (green) and N = 4 (blue) SLs on $\rm LaSrAlO_4$. Solid squares represent the conductivity obtained from the Drude model parameters in the legend of Fig. S7A, which exceeds the corresponding $\sigma^{dc}$ ($T=$ 175 K) values by less than 20 \%.
\label{dcS1}}
\end{figure}
We have used wide-band spectroscopic ellipsometry to accurately determine the dynamical electrical conductivity and permittivity of the SLs.
The distinct advantages of ellipsometry are as follows.
(i) In contrast to dc transport experiments, this method exposes the intrinsic electrodynamic response of the SLs, which is not influenced by the substrate, contacts, and extended defects. (ii) As a low-energy spectroscopic tool, it serves to determine critical parameters of the metal-insulator transition such as the energy gap and the density of carriers localized below $T_{MI}$. (iii) In comparison with other spectroscopic techniques, ellipsometry yields the frequency-dependent complex dielectric function without the need for reference measurements and Kramers-Kronig transformations. (iv) Variable angle ellipsometry is very sensitive to thin-film properties due to the oblique incidence of light, and it is generally used to derive optical constants of thin films and complex heterostructures \cite{Elli}.

The experimental setup comprises three ellipsometers to cover the spectral range of 12 meV to 6.5 eV. For the range 12 meV to 1 eV, we used a home-built ellipsometer attached to a standard Fast-Fourier-Transform Bruker 66v/S FTIR interferometer. The far-infrared measurements were performed at the infrared beamline IR1 of the Angström Quelle Karlsruhe ANKA synchrotron light source at the Karlsruhe Institute of Technology. For the mid-infrared measurements, we used the conventional glow-bar light source from a Bruker 66v/S FTIR. Finally, temperature dependencies of the pseudo-dielectric permittivity $\varepsilon^*_1$ at $\hbar \omega = 0.8$ eV were measured with a Woollam variable angle spectroscopic ellipsometer (VASE) equipped with an ultra high-vacuum cold-finger cryostat operated at $<5\times10^{-9}$ mbar chamber pressure. 

The inherent capacity of Woollam VASE ellipsometers to measure relative changes of the dielectric function on the order of $10^{-2}$ was boosted to an unprecedented level of $10^{-4}$ using temperature-modulation measurements of the dielectric constant at particular photon energies.
The ellipsometric angles $\Psi$ and $\Delta$ are defined through the complex Fresnel reflection coefficients for light polarized parallel ($r_p$) and perpendicular ($r_s$) to the plane of incidence, $\tan \Psi \ e^{i\Delta}=r_p/r_s$.
Figure S6 shows representative infrared spectra of $\Psi(\omega)$ and $\Delta(\omega)$ for the $N = 4$ and $2$ SLs and for the bare $\rm LaSrAlO_4$ and $\rm SrTiO_3$ substrates. The details of the data analysis have been discussed elsewhere \cite{CROCMO}. The SLs were treated as single-layer films according to an effective-medium approximation with a mixture of the nickelate and aluminate layers. A wavelength-by-wavelength regression procedure has been employed to extract the real and imaginary parts of the dielectric function \cite{VASE}. Figure S7  shows the best-fit model functions $\varepsilon_1(\omega)$ and $\varepsilon_2(\omega )$ obtained by inversion of the ellipsometric parameters in Fig. S6. The infrared spectra are well described by a broad Drude response $\varepsilon(\omega)= \varepsilon_\infty - \omega_{pl}^2/(\omega^2+i\omega\gamma)$ with a ratio of scattering rate and plasma frequency $\gamma / \omega_{pl} \approx 0.1-0.2$ that is typical for bulk complex oxides.
The parameters in the Drude fit are well constrained, because both $\varepsilon_1(\omega)$ and $\varepsilon_2(\omega)$ are available. The deviation of the Drude fit from the measured $\varepsilon_1(\omega)$ and $\varepsilon_2(\omega )$ below 30 meV in Fig. S7B (gray shadow area) can reflect the uncertainty in the inversion procedure for SLs on $\rm SrTiO_3$ due to the microwave dispersion of the ferroelectric soft mode of $\rm SrTiO_3$ \cite{Sirenko} and/or due to the presence of a dead layer with reduced conductivity at the substrate interface (Fig. S5B). This low-energy uncertainty does not, however, affect the relative spectral weight reduction,  $\Delta SW \approx 0.03\ (\pm 10 \%)$ per Ni atom within the gap energy range below $\Omega_G\approx 0.43$ eV, at the metal insulator transition in the $N = 2$ SL on $\rm SrTiO_3$.

The effective mass enhancement $m^{*}/m $ is estimated from the
plasma frequency as
\begin{equation}
m^{*}/m = \frac{4 \pi e^2 n}{m \omega_{pl}^2}\approx \frac{11.7}{(\omega_{pl},[eV])^2},
\end{equation}
where $n=\frac{1}{2}\times 1.7\times 10^{22}$ cm$^{-3}$, by assuming one electron per Ni atom.
We note that $\omega_{pl}$ is almost independent of $N$, implying the volume fraction of the metallic $\rm LaNiO_3$ layers remains the same in all SLs. With $\omega_{pl} \approx 1.1$ eV, as derived from the Drude model fit in Fig. 2A, we obtain $m^{*}/m \approx 10$ which is in good agreement with the value for bulk $\rm LaNiO_3$ from the specific heat measurements \cite{Xu}. Using the Fermi energy $E_F=0.5$ eV  derived from the thermopower of $\rm LaNiO_3$ \cite{Xu}, we estimate the Fermi velocity as
\begin{equation}
v_F = c \sqrt{\frac{2E_F}{m c ^2}\frac{m}{m^*}}\approx 1.33\times 10^{7} {\rm cm/s}  .
\end{equation}
The mean free path, $l$, can be estimated from
\begin{equation}
l\ {\rm [\AA]} = v_F\tau=\frac{v_F}{2\pi c\gamma}= 6.57\times 10^{-5}\frac{v_F \ {\rm [cm/s]}}{\gamma \ {\rm [meV]}}\approx \frac{874}{\gamma \ {\rm [meV]}}.
\end{equation}
With $\gamma\approx $200 meV (90 meV), as derived from the Drude model fit in Fig. 2A and Fig. S7A, we obtain $l =$ 4.4 \AA \ (9.7 \AA) for the $N=2$ ($N=4$) SL on $\rm LaSrAlO_4$. For the $N=2$ ($N=4$) SL on $\rm SrTiO_3$ we estimate $l =$ 6.4 \AA \ (12 \AA), respectively. Remarkably, the mean free path correlates with the individual $\rm LaNiO_3$ layer thickness, testifying, along with the constant volume fraction of the metallic layers, to  the atomic quality of the interfaces .

Our results indicate that, even in the $N=2$ samples at $T \gtrsim T_{MI}$, the conductivity of the $\rm LaNiO_3$ layers exhibits a clearly metallic temperature and frequency dependence. We define $T_{MI}$ as the temperature at which the temperature derivatives of both $\varepsilon_2(T)$ (Figs. 2A and 2B) and $\varepsilon_1(T)$ (Figs. 2C  and 2D) change sign. The consistent temperature evolution of $\varepsilon_1$ and $\varepsilon_2$ over a broad range of photon energies
demonstrates the intrinsic nature of the charge-localization transition observed in SLs with $N =2$. In the $\omega \rightarrow 0$ limit this criterion is analogous to a sign change of the temperature derivative of the $dc$ resistivity, $d\rho/dT$, observed at $T_{MI}$ in bulk $R$NiO$_3$. This is in contrast to results of recent $dc$ electrical resistivity measurements where the insulating behavior of 2 u.c. thick  $\rm LaNiO_3$ is attributed to variable range hopping transport \cite{May2} or film-substrate interface effects \renewcommand{\cite}[1]{S\citenum{#1}} (\cite{Scherwitzl}, \cite{Son}).\renewcommand{\cite}[1]{(S\citenum{#1})} 
For a thickness of $N=2 $ u.c. the behavior of $\rm (LaNiO_3)_N/(SrMnO_3)_2$ SLs is insulating over the entire temperature range, whereas the $N=4 $ u.c. SL is metallic with an upturn in resistivity below 50 K. Even in the $N=4$ u.c.  metallic sample, the mean free path $\l$ is  estimated to be less than a single unit cell \cite{May2}. This suggests that Anderson localization induced by disorder is responsible for the insulating behavior in these systems, in contrast to the sharp temperature dependence observed in our SLs that indicates a metal-insulator transition driven by collective interactions. Ultrathin single films of $\rm LaNiO_3$ show a crossover from metallic to insulating behavior at a larger thickness
\renewcommand{\cite}[1]{S\citenum{#1}} (\cite{Scherwitzl}, \cite{Son}),\renewcommand{\cite}[1]{(S\citenum{#1})}  which varies from 6 u.c. to 13 u.c. depending on the substrate. We argue that the presence of planar stacking fault defects and a dead layer with reduced conductivity at the substrate interface, as discussed above (Fig. S5), makes the analysis of the temperature-dependent resistivity measurements challenging and inconclusive about the conduction mechanism of ultrathin $\rm LaNiO_3$ films. Nevertheless, in order to directly compare our results with those reported in Refs.
\renewcommand{\cite}[1]{S\citenum{#1}} \cite{May2}, \cite{Scherwitzl}, and \cite{Son},\renewcommand{\cite}[1]{(S\citenum{#1})}
we have also performed $dc$ resistivity measurements on the $N=2$ and $N=4$ SLs on $\rm LaSrAlO_4$. Figure S8 shows that the  $dc$ resistivity of the $N=2$ SL exhibits a crossover from metallic  to insulating phase behavior below $T_{MI}\approx $ 150 K . The sharp temperature dependence in the insulating state does not fit to the stretched exponential function   \cite{May2} and can not be attributed to  variable range hopping transport. Figure S8 also shows that our $dc$ and optical conductivity data (Figs. 2A and S7A) are in close agreement. This resemblance once more indicates a low density of stacking faults in our samples.  

\paragraph*{Low-energy $\mu$SR instrumentation and data analysis\\}

Low energy muon spin rotation/relaxation (LE-$\mu $SR) uses $\sim 100 \%$ spin polarized positive muons of tunable keV-scale energy to study local magnetic properties of thin films or heterostructures as a function of the muon implantation depth. The details of the data acquisition and analysis have been described elsewhere \renewcommand{\cite}[1]{S\citenum{#1}} (\cite{Thomas}, \cite{muSR})
\renewcommand{\cite}[1]{(S\citenum{#1})}. More details of the LE-$\mu $SR methods and apparatus can be found on the website of the LEM group at Paul Scherrer Institute \cite{LEMweb}. This technique has been recently successfully applied to the case of magnetic ultra-thin films \cite{LEMmuSR1} and wires \cite{LEMmuSR2}.
\begin{SCfigure}[][htb]
\includegraphics*[width=0.56\columnwidth]{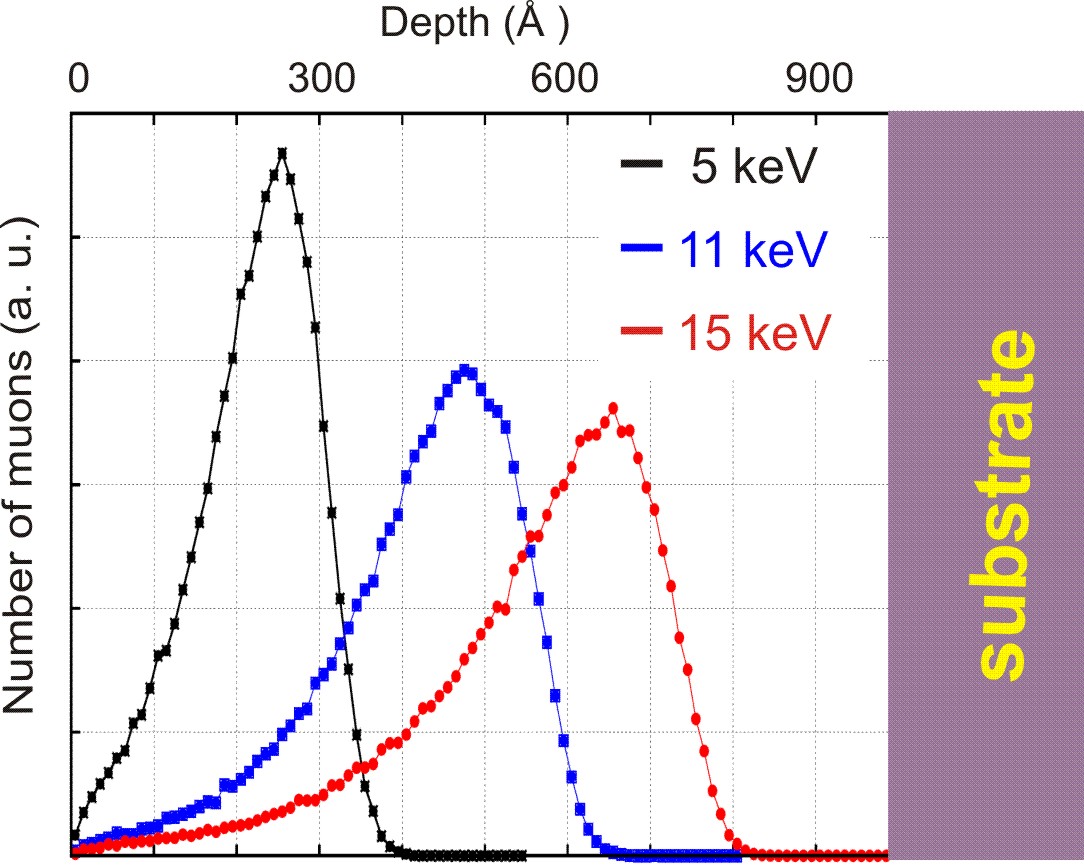}
\caption{Muon stopping profile in the $N=2$ SL showing the calculated probability that $\mu^+$ with an implantation energy of 5 keV (black), 11 keV (blue), and 15 keV (red) comes to rest at a certain depth near the surface.
\label{LEM1}}
\end{SCfigure}
\begin{figure}[htb]
\includegraphics*[width=\columnwidth]{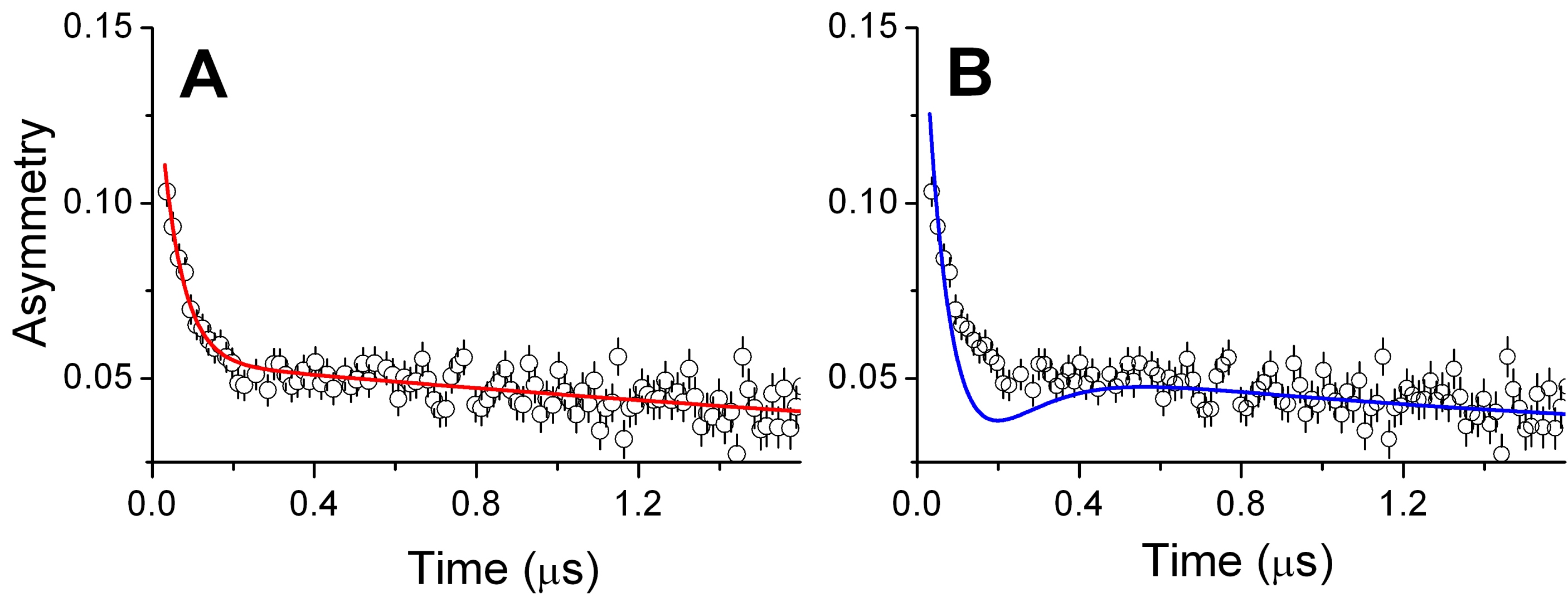}
\caption{Zero-field $\mu$SR function observed in the $N=2$ SL on $\rm LaSrAlO_4$ at 5 K. The solid lines represent the best-fit curves for (A) the two-component model function described in the manuscript and (B) Eq.(4), respectively.
\label{LEM2}}
\end{figure}
Figure S9 shows the muon stopping profile calculated for $\rm LaAlO_3| LaNiO_3$ SLs using the Monte Carlo algorithm TRIM.SP
\renewcommand{\cite}[1]{S\citenum{#1}} (\cite{Morenzoni}, \cite{Eckstein})
\renewcommand{\cite}[1]{(S\citenum{#1})}.

In our study we found that varying the stopping distribution of $\mu^+$ on the scale of about 50-800 $\AA$ through the control of the implantation energy between 5-15 keV  had no effect on the $\mu $SR spectra. The experimental LE-$\mu $SR curves in Figs. 3A to 3D were measured with muons of energy 10 keV, which are implanted at a mean depth of 45 nm. The initial asymmetry, $A(0)\approx 0.18$, is smaller than the asymmetry of the LE-$\mu $SR setup of $\approx 0.27$, because only 2/3 of the muon beam with a diameter of about 2 cm hit the sample with an area of $1 \times 2$ cm$^2$. The sample was surrounded by a Ni-coated sample holder, which causes a very fast depolarization ($< 0.06$ $\mu$s) of muons missing the sample.

The obtained spectra $\mu $SR spectra yield the probability distribution of the local magnetic field at the muon sites. As a local probe, $\mu$SR does not allow definite conclusions about the magnetic ordering pattern in the $N=2$ SLs. However, we rule out ferromagnetism based on an estimate of the ordered moment on the Ni sites from the $\mu $SR lineshape, $\mu_{Ni}\gtrsim 0.5\mu_B$ (see the main text of the manuscript). If these moment were co-aligned in the ordered state, the corresponding total moment $M=\mu_{Ni}n_{\rm Ni}V_{SL} \gtrsim 7.7\times 10^{-4}$ emu would have been readily detected in magnetization measurements. The absence of such an effect, which we confirmed in magnetometric measurements with sensitivity $\sim 10^{-7}$ emu.

We can also rule out a spin-glass state as the ground state of $N=2$ SLs, bearing in mind that oxygen deficient $\rm LaNiO_{2.75}$ exhibits spin-glass like behavior at low temperatures \cite{Sanchez}. A spin-glass state develops gradually due to randomly fluctuating local moments. In this case, the spin relaxation function should be exponential with a unique rate already at temperatures above about four times the actual glass transition temperature, $\sim $ 80 - 100 K \cite{Campbell}, which is at variance with the sharp temperature onset of the local moment observed in our data (Fig. 3A and solid squares in Fig. 4 of the manuscript). A similarly sharp transition was very recently observed by x-ray magnetic circular dichroism (XMCD) measurements in a magnetic field of 5T on a sample with $N=2$ grown under the same conditions \cite{XMCD}.

Additional evidence against a spin glass state can be derived from an analysis of the muon relaxation function. At low temperature, the spin-glass relaxation function in zero field can be described by \cite{Uemura}
\begin{equation}
A(t) = A_0[\frac{1}{3}\exp(-\sqrt{\lambda_dt})+\frac{2}{3}(1-\frac{\sigma^2t^2}{\sqrt{\lambda_dt+\sigma^2t^2}})\exp(-\sqrt{\lambda_dt+\sigma^2t^2}]
\end{equation}
with $\sigma\equiv\sqrt{q}\sigma_s$ and $\lambda_d\equiv 4\sigma_s^2(1-q)/\nu$, where $q$ is the Edwards-Anderson order parameter with the purely static and dynamic limits, $q =$ 1 and $q =$ 0, respectively, $\sigma_s$ is the static width of local fields at the muon site, and $\nu $ is the rate of the randomly fluctuating moments. This form of the relaxation function is expected for $\mu^+$ in  coexisting static and dynamic random local fields.
The fit of Eq. (4) to the time evolution of the zero-field muon spin polarization for the $N=2$ SL on $\rm LaSrAlO_4$ at 5 K (Fig. S10) gives reasonable parameters, i.e. close to the static limit with $q \approx$ 0.993, $\sigma_s = 11 - 15\ \mu $s$^{-1}$, and $\nu \approx 2\ \mu $s$^{-1}$. Nevertheless,
the simpler two-component model function, as described in the manuscript, provides a better fit to the data below 0.2 $\mu$s than the spin-glass function of Eq. (2). The analysis is consistent with long-range static antiferromagnetic order and confirms the conclusion of our manuscript.
